\documentclass[twocolumn,aps,prl,showpacs,floatfix]{revtex4}
\usepackage{graphicx}
\usepackage{times}
\usepackage{nicefrac}
\usepackage{amsmath}
\usepackage{amsfonts}
\usepackage{amssymb}
\usepackage{amsthm}
\usepackage{epsf}
\usepackage{bm}
\usepackage{bbm}
\usepackage{times}

\usepackage{dcolumn}

\newcommand{\balpha}{{\mbox{\boldmath$\alpha$}}}

\newcommand{\be}{\begin{eqnarray}}
\newcommand{\ee}{\end{eqnarray}}
\newcommand{\la}{\langle}
\newcommand{\ra}{\rangle}

\newcommand{\veps}{\varepsilon}
\newcommand{\vare}{\varepsilon}

\begin{document}

\title{QED corrections to the parity-nonconserving 6$s$-7$s$ amplitude
in $^{133}$Cs}

\author{V. M. Shabaev,$^{1}$ K. Pachucki,$^{2}$
I. I. Tupitsyn,$^{1}$  and V. A. Yerokhin$^{1,3}$}

\affiliation {$^{1}$Dept. of Physics, St.Petersburg State University,
Oulianovskaya 1, Petrodvorets, St.Petersburg 198504, Russia\\ $^{2}$ Institute
of Theoretical Physics, Warsaw University, Ho\.za 69, 00-681, Warsaw, Poland
\\$^{3}$Center of Advanced Studies, St.Petersburg State Polytekhnical University,
Politekhnicheskaya 29,
St.Petersburg 195251,
Russia}

\begin{abstract}
The complete gauge-invariant set of the one-loop QED corrections to the
parity-nonconserving 6$s$-7$s$ amplitude in $^{133}$Cs is evaluated to all orders
in $\alpha Z$ using a local version of the 
Dirac-Hartree-Fock potential. The calculations are peformed
in both length and velocity gauges for the absorbed photon.
The total binding QED correction is found
to be -0.27(3)\%, which differs from previous
evaluations of this effect. 
The weak charge of $^{133}$Cs, derived using two most accurate
values of the vector transition polarizability $\beta$, is 
$ Q_W=-72.57(46)$ for  
 $\beta = 26.957(51) a_{\rm B}^3$ 
and
$Q_W=-73.09(54)$ for
$\beta=  27.15(11)a_{\rm B}^3 $.  The first value
deviates by $1.1\sigma$ from the prediction of the Standard Model,
while the second one is in perfect agreement with it.

\end{abstract}
\maketitle

Investigations of parity noncoservation (PNC) effects in atomic
systems play a prominent role in tests of the Standard Model (SM)
and impose constraints on physics beyond  it \cite{khr91,khr04}.
The 6$s$-7$s$ PNC amplitude in $^{133}$Cs 
\cite{bou74} 
remains one
of the most attractive subject for such investigations.
The measurement of this amplitude to a 0.3\% accuracy
\cite{wood97,ben99} has stimulated a reanalysis
of related theoretical
contributions.
First, it was found 
\cite{der00,koz01,dzu01,dzu02}
that the role of the Breit interaction had been
underestimated in previous evaluations of this effect 
 \cite{dzu89,blu00}. Then, it was pointed out
 \cite{sush01} 
that the QED corrections may be comparable 
with the Breit corrections. The numerical evaluation
of the vacuum-polarization (VP) correction 
 \cite{joh01}
led to a 0.4\% increase
of the 6$s$-7$s$ PNC
amplitude in $^{133}$Cs, which resulted
in  a  2.2$\sigma$ deviation
of the weak charge of $^{133}$Cs
from the SM prediction.
This has triggered a great interest to calculations
of the one-loop QED corrections to the PNC amplitude.

While the VP contribution can easily be evaluated
to a high accuracy
within the Uehling approximation,
the calculation of the self-energy (SE) contribution is a much more demanding
problem (here and below we imply that the SE term embraces all one-loop
vertex
diagrams as well). To zeroth order in $\alpha Z$, it was derived
in Refs. \cite{mar83,lynn94}.
This correction, whose relative value equals 
to $-\alpha/(2\pi)$, is commonly 
 included in the definition of the nuclear weak charge. 
The $\alpha Z$-dependent part 
of the SE correction to the PNC matrix element between $s$ and $p$ states
was evaluated in Refs. \cite{kuch02,mil02}.
These calculations, 
which are exact to first order in  $\alpha Z$ and partially include
higher-order binding effects, yield the  correction of
-0.9(1)\% \cite{kuch02,kuch03} and -0.85\% \cite{mil02}. 
This restored the agreement with SM.

Despite of the close agreement of
the results obtained in Refs. \cite{mil02,kuch03},
the  status of the QED correction to
PNC in $^{133}$Cs cannot be considered
as resolved
until a complete $\alpha Z$-dependence calculation of
the SE correction to the 6$s$-7$s$ transition amplitude
 is accomplished. 
The reasons for that are the 
following. First, in case of cesium $(Z=55)$ the parameter
$\alpha Z \approx 0.4$ is not small and, therefore, the higher-order
corrections can be significant. 
Second, because the calculations \cite{kuch02,mil02,kuch03}
are performed for the PNC matrix element only,
they do not include other SE diagrams 
which contribute to the 6$s$-7$s$ transition amplitude.
For instance, these calculations do not account for diagrams 
in which the virtual photon
embraces both the weak interaction
and the absorbed photon. 
Our calculations, however, show that the contributions of all diagrams
are of the same order of magnitude (in both length and velocity
gauges, see below), and the final result arises through
a delicate cancellation of individual terms, none of which can be
neglected. 
Third, strictly speaking,
the PNC matrix element between the 
states of different  energies is not gauge invariant.
Despite the gauge-dependent part is suppressed by the small
energy difference \cite{mil02}, estimates of the uncertainty
in the definition of the PNC diagrams may fail due to
unphysical origin of the gauge-dependent terms.

The first step towards a complete $\alpha Z$-dependence calculation
was done in Ref. \cite{sap03}, where
the SE correction to the $2s$-$2p_{1/2}$ PNC transition in H-like ions
was evaluated. This transition was chosen to deal with 
the simplified gauge-invariant amplitude.
The results of that work agree with those of Refs.
\cite{kuch02,mil02,kuch03}. 
However, as was stressed there,
no claims can be made about the applicability of these
results to the 6$s$-7$s$ PNC transition in neutral cesium.
In this Letter we calculate
the whole gauge-invariant set of the one-loop QED corrections to 
the 6$s$-7$s$ PNC transition amplitude in $^{133}$Cs and compare
the obtained result with the previous treatments.

A systematic derivation of the QED corrections in a fully relativistic
approach requires the use of perturbation theory starting with a one-electron
approximation in an effective local potential $V(r)$. 
In neutral atoms, it is natural to assume that $V(r)$ includes 
not only the Coulomb field of the nucleus but also
a part of the electron-electron interaction.
 The  interaction
of the electrons with the quantized electromagnetic field and the correlation
effects 
are  accounted for by perturbation theory. 
In this way we obtain quantum electrodynamics in the Furry picture.

To derive formal expressions for the transition amplitude
we employ the method developed
in Ref. \cite{sha90} and described in detail in Ref. \cite{sha02}.
Since the wave length of the absorbed photon is much larger than
the atomic size, one can use the dipole approximation.
Within this approximation, calculations in the velocity
gauge are performed using formulas given 
in Ref. \cite{sha02} with the replacement $\exp{(i{\bf k}\cdot {\bf
    x})}\rightarrow 1$ in the photon wave function. The
corresponding formulas in the length gauge are easily obtained by 
 replacing $\balpha$ with ${\bf r}$ 
in all vertices
corresponding to the photon absorbtion and by multiplying
the  amplitude with the factor $i(E_b-E_a)$, where 
 $E_a$ and $E_b$ are the total energies
of the atom in the initial ($6s$) and final ($7s$) states, respectively.
This simple rule can
 be derived using Eq. (205) of Ref. \cite{sha02}
and the equal-time 
commutation relations.

\begin{widetext}
\begin{figure}
\begin{minipage}{16cm}
\centering
\includegraphics[clip=true,width=1.0\textwidth]{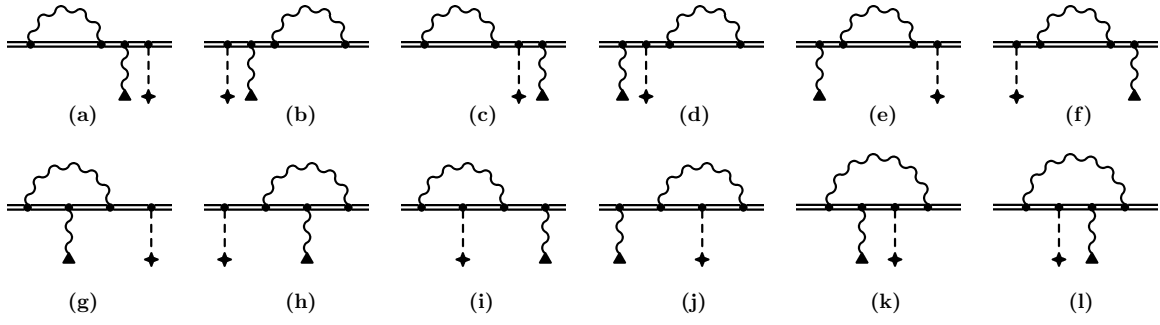}
\caption{Feynman diagrams for the SE corrections to
the PNC transition amplitude. The wavy line terminated
with a triangle indicates the absorbed photon. The dashed 
line terminated with a cross indicates the electron-nucleus weak interaction.
 \label{fig:sepnc}}
\end{minipage}
\end{figure}
\end{widetext}

To zeroth order, the 6$s$-7$s$ PNC transition amplitude, 
which is usually employed in these
calculations, is
\begin{eqnarray}\label{zero}
E_{\rm PNC} = \sum_n\Bigl[\frac{\la b|d_z|n\ra \la n|H_W|a\ra}{\vare_a-\vare_n}
+\frac{\la b|H_W|n\ra \la n|d_z|a\ra}{\vare_b-\vare_n}\Bigr]\,.
\end{eqnarray}
Here $a$ and $b$ denote the $6s$ and $7s$ one-electron states, respectively,
 with the angular momentum projections $m_a=m_b=1/2$,
$d_z=ez$ is the $z$ projection of the dipole moment operator ($e<0$),
$H_W=-(G_F/\sqrt{8})Q_W \rho_{\rm nuc}(r)\gamma_5 $
is the nuclear spin-independent weak-interaction Hamiltonian  \cite{khr91},
$G_F$ is the Fermi constant, $\gamma_5$ is the Dirac matrix, and
$\rho_{\rm nuc}$  is the weak-charge distribution. 
The one-loop SE corrections
are defined by diagrams presented in Fig.~\ref{fig:sepnc}.
The derivation of the formulas for these diagrams is very
similar to that for the  QED corrections to the transition
amplitude described in detail in Ref. \cite{sha02}. 
As a result, the SE correction
 is given by the sum of the following terms:
\begin{eqnarray}\label{se_a}
\delta E_{\rm PNC}^{\rm a}&=& \sum_{n_1,n_2}^{(n_1\ne b)}
\frac{\la b|\Sigma(\vare_b)|n_1\ra \la n_1|d_z|n_2\ra \la n_2 | H_W|a\ra}
{(\vare_b-\vare_{n_1}) (\vare_a-\vare_{n_2})}\nonumber\\
&&+\frac{1}{2}\sum_{n}
\frac{\la b|\Sigma'(\vare_b)|b\ra \la b|d_z|n\ra \la n | H_W|a\ra}
{ (\vare_a-\vare_{n})}\,,
\end{eqnarray}
\begin{eqnarray}\label{se_b}
\delta E_{\rm PNC}^{\rm b}&=& \sum_{n_1,n_2}^{(n_2\ne a)}
\frac{\la b|H_W|n_1\ra \la n_1|d_z|n_2\ra \la n_2 |\Sigma(\vare_a)|a\ra}
{(\vare_b-\vare_{n_1}) (\vare_a-\vare_{n_2})}\nonumber\\
&&+\frac{1}{2}\sum_{n}
\frac{\la b|H_W|n\ra \la n|d_z|a\ra \la a | \Sigma'(\vare_a)
|a\ra}{ (\vare_b-\vare_{n})}\,,
\end{eqnarray}
\begin{eqnarray}\label{se_c}
\delta E_{\rm PNC}^{\rm c}&=& \sum_{n_1,n_2}^{(n_1\ne b)}
\frac{\la b|\Sigma(\vare_b)|n_1\ra \la n_1|H_W|n_2\ra \la n_2 | d_z|a\ra}
{(\vare_b-\vare_{n_1}) (\vare_b-\vare_{n_2})}\nonumber\\
&&+\frac{1}{2}\sum_{n}
\frac{\la b|\Sigma'(\vare_b)|b\ra \la b|H_W|n\ra \la n | d_z|a\ra}
{ (\vare_b-\vare_{n})}\nonumber\\
&&-\sum_{n}
\frac{\la b|\Sigma(\vare_b)|b\ra \la b|H_W|n\ra \la n | d_z|a\ra}
{ (\vare_b-\vare_{n})^2}\,,
\end{eqnarray}
\begin{eqnarray}\label{se_d}
\delta E_{\rm PNC}^{\rm d}&=& \sum_{n_1,n_2}^{(n_2\ne a)}
\frac{\la b|d_z|n_1\ra \la n_1|H_W|n_2\ra \la n_2 |\Sigma(\vare_a)|a\ra}
{(\vare_a-\vare_{n_1}) (\vare_a-\vare_{n_2})}\nonumber\\
&&+\frac{1}{2}\sum_{n}
\frac{\la b|d_z|n\ra \la n|H_W|a\ra \la a | \Sigma'(\vare_a)
|a\ra}{ (\vare_a-\vare_{n})}\nonumber\\
&&-\sum_{n}
\frac{\la b|d_z|n\ra \la n|H_W|a\ra \la a | \Sigma(\vare_a)
|a\ra}{ (\vare_a-\vare_{n})^2}\,,
\end{eqnarray}
\begin{eqnarray} \label{se_e}
\delta E_{\rm PNC}^{\rm e}= \sum_{n_1,n_2}
\frac{\la b|d_z|n_1\ra \la n_1| \Sigma(\vare_a)|n_2\ra \la n_2 | H_W|a\ra}
{(\vare_a-\vare_{n_1}) (\vare_a-\vare_{n_2})}\,,
\end{eqnarray}
\begin{eqnarray}\label{se_f}
\delta E_{\rm PNC}^{\rm f}= \sum_{n_1,n_2}
\frac{\la b|H_W|n_1\ra \la n_1| \Sigma(\vare_b)|n_2\ra \la n_2 | d_z|a\ra}
{(\vare_b-\vare_{n_1}) (\vare_b-\vare_{n_2})}\,,
\end{eqnarray}
\begin{eqnarray}\label{se_g}
\delta E_{\rm PNC}^{\rm g}&=&\frac{i}{2\pi} \int_{-\infty}^{\infty} d\omega
\sum_{n,n_1,n_2}
\frac{\la n_1|d_z|n_2\ra
\la n | H_W|a\ra}{(\vare_a-\vare_{n})}\nonumber\\
&&\times
\frac{\la b n_2|I(\omega)|n_1 n\ra}
{[\vare_b-\omega-u\vare_{n_1}]
 [\vare_a-\omega-u\vare_{n_2}]}\,,
\end{eqnarray}
\begin{eqnarray}\label{se_h}
\delta E_{\rm PNC}^{\rm h}&=&\frac{i}{2\pi} \int_{-\infty}^{\infty} d\omega
\sum_{n,n_1,n_2}
\frac{\la b | H_W|n\ra 
\la n_1|d_z|n_2\ra}{(\vare_b-\vare_{n})}\nonumber\\
&&\times \frac{\la n n_2|I(\omega)|n_1 a\ra}
{ [\vare_b-\omega-u\vare_{n_1}]
 [\vare_a-\omega-u\vare_{n_2}]}\,,
\end{eqnarray}
\begin{eqnarray}\label{se_i}
\delta E_{\rm PNC}^{\rm i}&=&\frac{i}{2\pi} \int_{-\infty}^{\infty} d\omega
\sum_{n,n_1,n_2}
\frac{\la n_1|H_W|n_2\ra
\la n | d_z|a\ra}{(\vare_b-\vare_{n})}\nonumber\\
&&\times\frac{\la b n_2|I(\omega)|n_1 n\ra}
{[\vare_b-\omega-u\vare_{n_1}]
 [\vare_b-\omega-u\vare_{n_2}]}\,,
\end{eqnarray}
\begin{eqnarray}\label{se_j}
\delta E_{\rm PNC}^{\rm j}&=&\frac{i}{2\pi} \int_{-\infty}^{\infty} d\omega
\sum_{n,n_1,n_2}
\frac{\la b | d_z|n\ra 
\la n_1|H_W|n_2\ra}{(\vare_a-\vare_{n})}
\nonumber\\
&&\times \frac{\la n n_2|I(\omega)|n_1 a\ra}
{ [\vare_a-\omega-u\vare_{n_1}]
[\vare_a-\omega-u\vare_{n_2}]}\,,
\end{eqnarray}
\begin{eqnarray}\label{se_k}
\delta E_{\rm PNC}^{\rm k}&=&\frac{i}{2\pi} \int_{-\infty}^{\infty} d\omega
\sum_{n_1,n_2,n_3}
\frac{\la b n_2|I(\omega)|n_1 a\ra}{[\vare_b-\omega-u\vare_{n_1}]}
\nonumber\\
&&\times\frac{\la n_1|d_z|n_3\ra
\la n_3 | H_W|n_2\ra}
{[\vare_a-\omega-u\vare_{n_3}]
 [\vare_a-\omega-u\vare_{n_2}]}\,,
\end{eqnarray}
\begin{eqnarray}\label{se_l}
\delta E_{\rm PNC}^{\rm l}&=&\frac{i}{2\pi} \int_{-\infty}^{\infty} d\omega
\sum_{n_1,n_2,n_3}
\frac{\la b n_2|I(\omega)|n_1 a\ra}{[\vare_b-\omega-u\vare_{n_1}]}\nonumber\\
&&\times\frac{ \la n_1|H_W|n_3\ra
\la n_3 | d_z|n_2\ra}
{[\vare_b-\omega-u\vare_{n_3}]
 [\vare_a-\omega-u\vare_{n_2}]}\,.
\end{eqnarray}
Here the SE operator is defined by
\begin{eqnarray}
\la c|\Sigma(E)|d\ra  \equiv  \frac{i}{2\pi} \int_{-\infty}^{\infty}
d\omega \;
\sum_{n}\frac{\la cn|
I(\omega)|n d\ra}{E-\omega-u\veps_n} \,,
\end{eqnarray}
$I(\omega) \equiv  e^2\alpha^{\mu}\alpha^{\nu}D_{\mu \nu}(\omega)$,
$\alpha^{\mu}\equiv \gamma^0\gamma^{\mu}=(1,\balpha)$, 
$D_{\mu \nu}(\omega)$ is the photon propagator defined as
in Ref. \cite{sha02}, $\Sigma'(E)=d\Sigma(E)/dE$,
and $u=1-i0$ ensures the correct position of poles of the electron
propagators with respect to the integration contour.
Taking into account the corresponding diagrams with the mass counterterm 
results in the replacement $\Sigma(E)\rightarrow
\Sigma(E)-\gamma^0\delta m$.
The  expressions for the VP corrections, which
do not contain any insertions with the external photon line or the
weak interaction attached to the electron loop,  
are obtained from
 Eqs. (\ref{se_a})-(\ref{se_f}) by the replacement of the SE operator with
the VP potential. The 
other VP corrections
will not be considered  here, since their contribution
is negligible.

The corresponding expressions in the velocity gauge are 
obtained by the replacement  $d_z \rightarrow -ie\alpha_z/(E_b-E_a)$,
where the energies $E_a$ and $E_b$ include the QED corrections.
In addition to the replacement
$d_z \rightarrow -ie\alpha_z/(\vare_b-\vare_a)$
in Eqs. (\ref{zero})-(\ref{se_l}), it yields the contribution 
\begin{eqnarray} \label{se_add}
\delta E_{\rm PNC}^{\rm add}=
-\frac{\la b|\Sigma(\vare_b)|b\ra - \la a|\Sigma(\vare_a)|a\ra}
{\vare_b-\vare_a}E_{\rm PNC}\,,
\end{eqnarray}
which results from the expansion of the factor $1/(E_b-E_a)$.

Formulas (\ref{se_a})-(\ref{se_add})
 contain ultraviolet and infrared divergences.
To cancel the ultraviolet divergences, we 
expand 
 contributions (\ref{se_a})-(\ref{se_f}) into
zero-, one-, and many-potential terms
and  contributions (\ref{se_g})-(\ref{se_j})
into zero- and many-potential terms.
The ultraviolet divergencies are present only in the
zero- and one-potential terms. They are removed
analytically by calculating these terms in the momentum space
 (for details, we refer to Refs. \cite{sny91,yer99,sap02}). 
The many-potential terms are evaluated in configuration space.
The infrared divergences, which occur in contributions 
(\ref{se_a})-(\ref{se_d}) and (\ref{se_k})-(\ref{se_l}),
are regularized by introducing a nonzero photon mass and cancelled
analytically.

Since the levels $6s$, $6p_{1/2}$, $7s$, and $7p_{1/2}$
are very close to each other, to get reliable results for the
transition amplitude under consideration,
one needs to use a local potential $V(r)$ that reproduces
 energies and wave functions of these states to a 
sufficient accuracy. 
We construct such a potential by inverting the radial Dirac equation
with the radial wave function obtained by solving the 
Dirac-Hartree-Fock (DHF) equation
with the code of Ref. \cite{bra77}.
Details of this procedure will be published elsewhere.
In Table 1, we compare the energies obtained with the local potential
$V(r)$, that was derived using mainly the DHF wave function of the $6s$ state,
with the DHF energies and with the experimental ones.

\begin{table}
\caption{The binding energies of low-lying states in Cs, in a.u.} \vspace{0.2cm}
\begin{ruledtabular}
\begin{tabular} {crrr}
State &\multicolumn{1}{c}{Loc. pot.}
                 & \multicolumn{1}{c}{DHF}
                                     & \multicolumn{1}{c}{Exp.} \\
\hline
$6s_{1/2}$ & -0.13079 & -0.12824  & -0.14310 \hspace*{0.115cm}
\\
$6p_{1/2}$ & -0.08696 &  -0.08582 & -0.09217 \hspace*{0.115cm}
\\
$7s_{1/2}$ & -0.05621   & -0.05537 & -0.05865 \hspace*{0.115cm}
\\
$7p_{1/2}$ & -0.04251 &  -0.04209 & -0.04393 \hspace*{0.115cm}
\end{tabular}
\end{ruledtabular}
\end{table}

Numerical evaluation of expressions (\ref{zero})-(\ref{se_add})
was performed by employing the dual-kinetic-balance 
finite basis set method \cite{sha04} with basis
functions construced from B-splines. 
The calculation of
the zeroth-order contribution, with $V(r)$
constructed as indicated above, 
yields $E_{\rm PNC}=$-1.002,
in units i$\times 10^{-11}Q_W/(-N)$ a.u. This value should be
compared with the corresponding DHF value, -0.742 \cite{koz01}, and with the
 value that includes the correlation effects, 
-0.908 \cite{dzu89}. The results for the SE corrections
are presented in Table 2.
Since there is a significant cancellation between terms containing
the infrared singularities, the terms corresponding to
$n=a$ in  $\Sigma'(\vare_a)$ and $n=b$ in  $\Sigma'(\vare_a)$
are subtracted from contributions (\ref{se_a})-(\ref{se_d})  and added to 
 contributions  (\ref{se_k})-(\ref{se_l}).
The total SE correction 
$\delta E_{\rm PNC}^{\rm tot}$ contains also the free term,
$-\alpha/(2\pi)E_{\rm PNC}$, mentioned above. Since this term is usually
included into the weak charge $Q_W$, one has to consider
the binding SE correction defined as 
$\delta E_{\rm PNC}^{\rm bind}=\delta E_{\rm PNC}^{\rm tot}
+\alpha/(2\pi)E_{\rm PNC}$. According to Table 2,
the binding SE correction amounts to -0.67\%.
To estimate the uncertainty
of this value due to  correlation effects,
  we have also performed the calculations 
with $V(r)$ constructed employng the DHF wave function of the
$7s$ state. While this leads to a 
 2\% decrease of the transition amplitude,  the 
relative shift of the SE correction is five times smaller.
Since the correlation effects contribute to the transition amplitude
on the 20\% level,
we assume a 4\% uncertainty for the total SE correction. 
Therefore, our value for the binding SE correction
is -0.67(3)\%.
This value differs from the previous evaluations of the SE effect,
-0.9(1)\% \cite{kuch03} and -0.85\% \cite{mil02}.

We have also calculated the VP correction.
Our value for the Uehling part amounts to 0.410\%, which agrees
well with the previous calculations of this effect.
We have found that including the screening into
the Uehling potential does not affect this value.
As to the Wichmann-Kroll (WK) correction, our calculation
employing approximate formulas for the WK potential from
Ref. \cite{fain91} yields  -0.004\% (cf. \cite{dzu02}). 
This leads to the 0.406\%
result for the total VP correction. Therefore, the total binding QED
correction amounts to -0.27(3)\%.

To get the total $6s$-$7s$ PNC transition amplitude in $^{133}$Cs,
we combine the value that includes the correlation
effects \cite{dzu89,blu00,koz01,dzu02}, -0.908(1)$\pm$ 0.5\%,
with the -0.61\% Breit correction \cite{dzu02},
the -0.27(3)\% binding QED correction, the -0.19(6)\%
neutron skin correction \cite{der02}, the -0.08\%
correction due to the renormalization of $Q_W$ from the
atomic momentum transfer $q\sim 30$ MeV down to $q=0$
\cite{mil02}, and the 0.04\% contribution from the 
electron-electron weak interaction \cite{mil02}. Using the experimental
value for the $E_{\rm PNC}/\beta$
\cite{wood97}, where $\beta$ is the
vector transition polarizabilty,  we obtain for the weak charge
of $^{133}$Cs:
\begin{eqnarray}
Q_W=-72.57(29)_{\rm exp}(36)_{\rm th}
\end{eqnarray}
for $\beta = 26.957(51) a_{\rm B}^3$ 
\cite{ben99,dzu02} and 
\begin{eqnarray}
Q_W=-73.09(39)_{\rm exp}(37)_{\rm th}
\end{eqnarray}
for $\beta= 27.15(11) a_{\rm B}^3$ 
\cite{cho97,vas02,dzu02}.
We conclude that  the first value
deviates from the SM prediction
 of -73.09(3) \cite{gro00}
 by 1.1.$\sigma$, while the second one
is in perfect agreement with it.

In summary, we have calculated the QED correction to the
$6s$-$7s$ PNC transition amplitude in $^{133}$Cs and derived
the weak charge using two most accurate values
of the vector transition polarizability.
Further improvement of atomic tests of the Standard Model can be achieved,
from theoretical side,  by more accurate calculations
of the electron-correlation effects and, from experimental side, by more
precise measurements of the PNC amplitude in cesium or other atomic systems.
Particularly interesting is the francium  atom, where PNC effects
are greatly enhanced by strong electric field of the nucleus.
Precise measurements of the PNC amplitude in Fr are becoming feasible due to
recent advances in producing, storing, and
cooling of short-lived radioactive isotopes.

This work was supported by NATO (Grant No. PST.CLG.979624), 
by EU (Grant No. HPRI-CT-2001-50034),
and by RFBR (Grant No. 04-02-17574).

\begin{table}
\caption{The SE corrections to the $6s-7s$ PNC amplitude in $^{133}$Cs,
in \%. The results are presented in both the length (L)
and the velocity (V) gauge.} \vspace{0.2cm} 
\begin{ruledtabular}
\begin{tabular} {crrcrr}
Contr. &\multicolumn{1}{c}{L-gauge}
                 & \multicolumn{1}{c}{V-gauge}
& Contr. &\multicolumn{1}{c}{L-gauge}
                 & \multicolumn{1}{c}{V-gauge}
                                     \\
\hline
$\delta E_{\rm PNC}^{\rm a}$ &-0.09 & -0.11 &
$\delta E_{\rm PNC}^{\rm h}$ & -4.04 & -3.40 \hspace*{0.115cm}
\\
$\delta E_{\rm PNC}^{\rm b}$ & 1.31 & 1.11 &
$\delta E_{\rm PNC}^{\rm i}$ & -4.61 & -3.97 \hspace*{0.115cm}
\\
$\delta E_{\rm PNC}^{\rm c}$ & 0.34 & 0.40 &
$\delta E_{\rm PNC}^{\rm j}$ & 1.49 & 1.73 \hspace*{0.115cm}
\\
$\delta E_{\rm PNC}^{\rm d}$ & -0.38 & -0.32 &
$\delta E_{\rm PNC}^{\rm k}$ & -0.79 & -1.03 \hspace*{0.115cm}
\\
$\delta E_{\rm PNC}^{\rm e}$ &-1.29 & -1.53 &
$\delta E_{\rm PNC}^{\rm l}$ & 2.05 & 1.41 \hspace*{0.115cm}
\\
$\delta E_{\rm PNC}^{\rm f}$ & 3.89 & 3.25 &
$\delta E_{\rm PNC}^{\rm add}$ & 0.00 & 0.10 \hspace*{0.115cm}
\\
$\delta E_{\rm PNC}^{\rm g}$ & 1.33 & 1.57 &
$\delta E_{\rm PNC}^{\rm tot}$ & -0.79 & -0.79 \hspace*{0.115cm}
\\
& & &
$\delta E_{\rm PNC}^{\rm bind}$ & -0.67 & -0.67  
\hspace*{0.115cm}
\end{tabular}
\end{ruledtabular}
\end{table}


\end{document}